\documentclass[conference]{IEEEtran}
\usepackage[T1]{fontenc}

\usepackage{cite}
\usepackage{amsmath,amssymb,amsfonts}
\usepackage{algorithmic}
\usepackage{stfloats}
\usepackage{graphicx}
\usepackage{textcomp}
\usepackage{xcolor}
\usepackage{todonotes}
\usepackage{subcaption}
\usepackage{printlen}  % find the text widths for creating figures
\usepackage[inline]{enumitem}
\usepackage{siunitx}
\PassOptionsToPackage{hyphens}{url} % fixes line breaking at URL footnotes
\usepackage[hidelinks]{hyperref}

\usepackage{cleveref}
\usepackage{tabularray}
\usepackage[most]{tcolorbox}
\usepackage{comment}

\crefformat{figure}{#2Fig.~#1#3}
\Crefformat{figure}{#2Figure~#1#3}

\DeclareSIUnit\hour{h}

\makeatletter
\newcommand\footnoteref[1]{\protected@xdef\@thefnmark{\ref{#1}}\@footnotemark}
\makeatother

\def\BibTeX{{\rm B\kern-.05em{\sc i\kern-.025em b}\kern-.08em
    T\kern-.1667em\lower.7ex\hbox{E}\kern-.125emX}}
\begin{document}

%\bstctlcite{IEEEexample:BSTcontrol} % force et al for more than 3 authors in bibliography

\title{Container-level Energy Observability in\\ Kubernetes Clusters
% \thanks{Identify applicable funding agency here. If none, delete this.}
}

\author{\IEEEauthorblockN{Bjorn Pijnacker}
\IEEEauthorblockA{\textit{University of Groningen}\\
Groningen, the Netherlands \\
b.pijnacker@rug.nl}
\and
\IEEEauthorblockN{Brian Setz}
\IEEEauthorblockA{\textit{University of Groningen}\\
Groningen, the Netherlands \\
b.setz@rug.nl}
\and
\IEEEauthorblockN{Vasilios Andrikopoulos}
\IEEEauthorblockA{\textit{University of Groningen}\\
Groningen, the Netherlands \\
v.andrikopoulos@rug.nl}
}

\maketitle

\begin{abstract}
Kubernetes has been for a number of years the default cloud orchestrator solution across multiple application and research domains.
As such, optimizing the energy efficiency of Kubernetes-deployed workloads is of primary interest towards controlling operational expenses by reducing energy consumption at data center level and allocated resources at application level.
A lot of research in this direction aims on reducing the total energy usage of Kubernetes clusters without establishing an understanding of their workloads, i.e. the applications deployed on the cluster.
This means that there are untapped potential improvements in energy efficiency that can be achieved through, for example, application refactoring or deployment optimization.
For all these cases a prerequisite is establishing fine-grained observability down to the level of individual containers and their power draw over time.
A state-of-the-art tool approved by the Cloud-Native Computing Foundation, Kepler, aims to provide this functionality, but has not been assessed for its accuracy and therefore fitness for purpose.
In this work we start by developing an experimental procedure to this goal, and we conclude that the reported energy usage metrics provided by Kepler are not at a satisfactory level.
As a reaction to this, we develop KubeWatt as an alternative to Kepler for specific use case scenarios, and demonstrate its higher accuracy through the same experimental procedure as we used for Kepler.
\end{abstract}


\section{Introduction} \label{sec:introduction}

Datacenters and cloud computing are significant power users. In 2021, cloud computing accounted for approximately \qty{1}{\percent} of global power usage, with it estimated to reach \qty{8}{\percent} before 2030\footnote{\url{https://spectrum.ieee.org/cloud-computings-coming-energy-crisis}}. As the leading container orchestration platform, Kubernetes-based workloads play a significant role in this power consumption. According to a 2022 Red Hat report, up to \qty{70}{\percent} of IT organizations use Kubernetes (K8s) in some way\footnote{\url{https://www.redhat.com/en/resources/state-of-enterprise-open-source-report-2022}}. In this respect, the energy usage and efficiency of applications running in Kubernetes clusters is of significant interest, since significant savings can be made for both cost and carbon emissions in data center-based computing.

Existing solutions attempt to optimize Kubernetes cluster or even data center power usage, and therefore energy consumption, as a whole \cite{douhara_kubernetes-based_2020,ghafouri_smart-kube_2023}. This has yielded some promising results. However, a relatively unexplored aspect of Kubernetes energy optimization is that of targeting energy usage at workload deployment level. 
To achieve any yield in this effort, it becomes essential to first establish \emph{energy observability} in K8s clusters across different levels of granularity: from complete clusters and all the way down to individual containers running in pods.
Having such observability features is also an essential building block towards addressing the challenge of efficient carbon footprint measurement as discussed by Jay et al.~\cite{jay2023}, especially given that K8s clusters often contain multiple workloads from potentially different tenants.

The current state-of-the-art energy measuring tool specifically designed for this purpose is Kepler~\cite{amaral_kepler_2023,amaral2024process}, which at the time of writing this paper is a sandbox maturity project under the Cloud Native Computing Foundation (CNCF) umbrella\footnote{https://www.cncf.io/projects/kepler/}. 
While Kepler has already been used in a number of research works as a source of energy consumption data~\cite{gudepu_demonstrating_2024,soldani_ebpf_2023,choochotkaew_advancing_2023}, beyond an initial evaluation in~\cite{amaral_kepler_2023} there has been no systematic evaluation of its accuracy in the literature --- at least to the extent of our knowledge.
This is particularly important because as indicated by Centofanti et al.~\cite{centofanti_impact_2024}, there are discrepancies observed in the reported measurements under experimental conditions for this tool. This follows a trend among tools with similar purposes~\cite{jay2023} since they rely on power modeling instead of actual measurements.

As such, this work aims to assess Kepler's fitness for purpose, and where necessary to provide alternatives and improvements.
In this effort, we adopt an empirical stance, and design a replicable experimental procedure which we execute under controlled conditions.
We collect and interpret the resulting data, and based on our findings we opt to develop an alternative to Kepler that does not appear to have the same issues for the same assessment.

The rest of this paper is as a result structured as follows. In the following section (\Cref{sec:kepler}), we present Kepler in more depth as a background for this study; in \Cref{sec:kepler-evaluation} we design and present an experimental evaluation of Kepler's accuracy. In \Cref{sec:kubewatt} we introduce KubeWatt as an alternative approach to Kepler, which we evaluate using the same procedure as we used for Kepler in \Cref{sec:kubewatt-evaluation}. Related research is discussed in \Cref{sec:related-work} and conclusions are drawn in \Cref{sec:conclusions}.

\section{CNCF Kepler} \label{sec:kepler}
The \textbf{K}ubernetes-based \textbf{E}fficient \textbf{P}ower \textbf{L}evel \textbf{E}xporte\textbf{r} (Kepler)~\cite{amaral_kepler_2023,amaral2024process} is a CNCF project that aims to estimate the power consumption of different Kubernetes components and export this data to Prometheus, a time-series database. At its core, Kepler uses the extended Berkeley Packet Filter (eBPF)\footnote{\url{https://ebpf.io/what-is-ebpf/}}, a technology to allow running programs in the Linux kernel to obtain energy-related system metrics. It also collects various real-time power consumption metrics using different sources, such as RAPL for CPU and DRAM, NVML for NVIDIA GPU power, the ACPI power management interface, Redfish or IPMI for platform power, or regression-based models when no real-time power metrics are available\footnote{\label{note:kepler-blog}\url{https://www.cncf.io/blog/2023/10/11/exploring-keplers-potentials-unveiling-cloud-application-power-consumption/}}.

By combining utilization metrics with platform and component power usage, Kepler can estimate the power consumption of each process, container or pod. This is done by dividing the total power consumption in \emph{idle} and \emph{dynamic power}, using Kepler's so called \emph{ratio power model}. Dynamic power is directly related to resource utilization and therefore attributed to the process it is responsible for the resource usage. The idle power of the host is then distributed among processes in accordance with their size, as is stipulated in the GreenHouse Gas protocol\footnote{\url{https://ghgprotocol.org/}} guidelines. 

A distinction is made between node metrics and container metrics. The node metrics are collected for each Kubernetes node and are split into core, DRAM, package, platform, and uncore components. The power data source that is used determines how exactly the power consumption of each component is derived, and multiple data sources are possible. For example, when using Redfish, the platform power is taken directly from Redfish while the other components are derived from RAPL. The container metrics on the other hand are collected for each container that is running on the Kubernetes node, and these include the same components that are collected on node level. 

Kepler can be deployed in two different configurations depending on the metrics that are available in the host environment. In its most basic deployment Kepler estimates the power consumption using utilization metrics of the host and feeding this data to a pre-trained power estimation model. It is also possible to train a custom power estimation model. The second deployment mode does not use a trained model, but instead retrieves power metrics directly from the bare-metal host using one of the aforementioned power metric sources. Kepler can also use a combination of the two configurations if some bare-metal metrics are missing, to estimate these missing components. A third deployment mode is hypothesized but not currently available. This mode enables Kepler to perform its calculations inside virtual machines. This requires Kepler to be deployed on the bare-metal host as well as on the virtual machine, monitoring the idle and dynamic power of each VM using the bare-metal metrics, and then expose this power data to the Kepler instance running inside the VM itself\footnoteref{note:kepler-blog}.

While Kepler is applied in several academic works, its accuracy and fitness for purpose have not, and to the extent of our knowledge, been evaluated yet. Gudepu et al.~\cite{gudepu_demonstrating_2024} use Kepler to obtain power measurements and indicate it produces similar results to Scaphandre, a tool which predicts power usage per resource. However, neither tool is validated using a source of ground truth power consumption. Additionally, the Kepler results are not published, it is only mentioned that the results are similar to Scaphandre. Soldani et al. use Kepler as a demonstration of an eBPF use-case. While they show a dashboard of energy measurements, these are not validated as being accurate~\cite{soldani_ebpf_2023}. Centofanti et al.~\cite{centofanti_impact_2024} investigate Kepler in comparison to Scaphandre and s-tui, a graphical stress-test and CPU monitoring tool\footnote{\url{https://github.com/amanusk/s-tui}}. However, for Kepler they only consider a configuration using the pre-trained power estimation model. They find that there are significant discrepancies between the tools used in their tests and conclude that further research is required to increase the robustness of these tools. Andringa \cite{andringa_estimating_2024} also investigates Kepler, concluding that Kepler's metrics are not accurate where total cluster power is concerned, but the discussed evaluation lacks the necessary depth and the per-container power attribution is not investigated further. 

It is clear that these previous works are not sufficient to validate Kepler's accuracy and fitness for purpose. To evaluate these properties, we must first validate that the power reported per container by Kepler matches closely to what is expected taking into account resource usage by that workload. Power draw measurements are fundamental towards calculating energy consumption at this level for each workload.

\section{Kepler Evaluation} \label{sec:kepler-evaluation}

We are interested in evaluating Kepler in terms of its main goal: container power attribution in Kubernetes clusters over time. To this end, the following research question is asked:

\begin{tcolorbox}[left=3pt, right=3pt, top=3pt, bottom=3pt]
\begin{enumerate}[label=\textbf{RQ\arabic*},left=0pt]
    \item How well does Kepler attribute power usage to containers on a Kubernetes node?
\end{enumerate}
\end{tcolorbox}

In order to answer this question empirically we design an experimental procedure that we discuss in the following, starting from the system under test.

\subsection{System under Test} \label{sec:kepler-evaluation:system-under-test}
\begin{figure}[t]
    \centering
    \includegraphics[width=\linewidth]{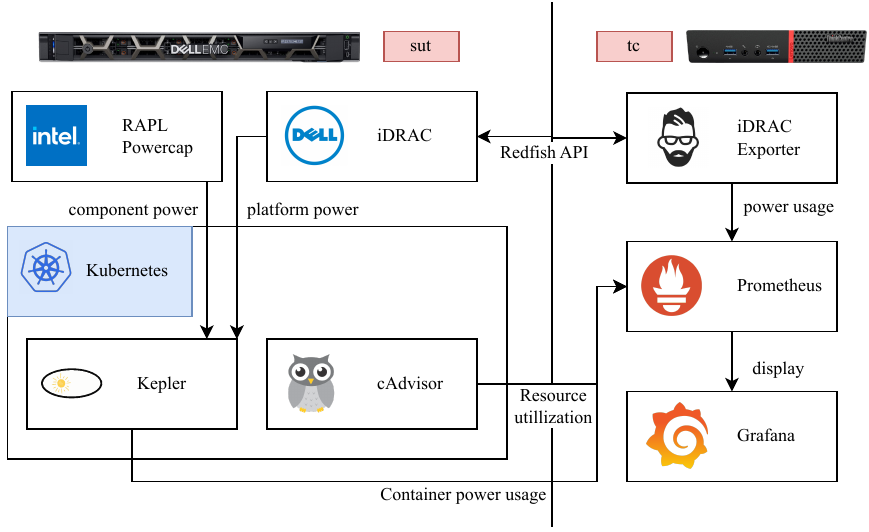}
    \caption{The hardware and software setup to evaluate Kepler. \texttt{sut} = system under test, \texttt{tc} = thin client.}
    \label{fig:kepler-evaluation:system-under-test:setup}
\end{figure}

Before Kepler's accuracy can be evaluated through experimentation, the system under test and its constituent software and hardware stacks need to be clearly defined. The critical components of the setup include a Kubernetes cluster to run workloads in, an observability stack to collect node and container level metrics, and a way to measure the ground-truth power consumption of the cluster. The complete setup for this purpose is shown in \cref{fig:kepler-evaluation:system-under-test:setup}.

\begin{table*}
    \caption{Metrics collected and used for the experiments}
    \label{tab:metrics}
    \centering
    \begin{tblr}{
            colspec={llX},
            width=.85\linewidth,
            column{2}={cmd=\detokenize,font=\ttfamily},
            row{1}={font=\bfseries},
        }
        \hline
        Source   & Metric                                        & Explanation                                                                                                 \\\hline
        iDRAC    & power_control_avg_consumed_watts              & Average power consumption in \si{\watt} over the last measured interval (\qty{1}{\minute})                                                                 \\
        cAdvisor & container_cpu_usage_seconds_total             & Cumulative CPU time consumed in seconds \\
        Kepler   & container_joules_total                        & Aggregated platform and component power per container in \si{\joule} \\
                 & container_cpu_instructions_total              & Number of CPU instructions measured per container \\
        %KubeWatt & kubewatt_container_power_watts                & Aggregated power per container in \si{\watt} \\
        \hline
    \end{tblr}
\end{table*}

The hardware setup that is used consists of two machines: a System Under Test (SUT), and a Thin Client (TC) for data collection, analysis, and visualization. The SUT is a Dell PowerEdge R640 server. It is equipped with two Intel Xeon Gold 6226R processors totaling 32 cores and 64 threads, \qty{96}{\giga\byte} of RAM and \qty{256}{\giga\byte} of RAID1 SSD storage. Fedora Server 40 is installed as the operating system. The server is also equipped with Dell iDRAC9 from which we can obtain, among others, the ground-truth power metrics using the Redfish API integration. We verified that the Redfish API data is indeed the ground-truth by including an external power monitoring wall plug.
The TC is a Lenovo ThinkCentre M910q with an Intel Core i3-6100T, \qty{8}{\giga\byte} of RAM, and \qty{256}{\giga\byte} of NVMe storage running Fedora Server 40 as the operating system. This machine is used to collect and analyze the data from the tests we perform on the SUT so that they do not influence test results of the SUT.

A single-node Kubernetes cluster is deployed on the SUT. This cluster is bootstrapped using Rancher Kubernetes Engine (RKE)\footnote{\url{https://github.com/rancher/rke}} version 1.5.8. RKE allows us to administer the cluster without running specific software on-node, as it uses SSH to set up the cluster on the node. For the purposes of the experiments, the exact Kubernetes distribution does not matter as long as it can be installed on bare-metal Linux. The cluster is set up in its most basic form, and we remove any workloads that are not necessary for our tests such as \texttt{nginx-ingress-controller} in order to reduce the amount of noise in our power and CPU metrics.

The backbone of the observability stack that is used to collect various metrics from the SUT consists of Prometheus and Grafana. Prometheus is the monitoring system and time series database, and Grafana is used for visualization. All components that produce metrics that require collecting support collection by Prometheus, and as Prometheus has CNCF graduated maturity, it is the recommended choice for this use case\footnote{\url{https://www.cncf.io/projects/prometheus/}}. Both Prometheus and Grafana are deployed and configured as services on the TC, so the data processing does not affect the power usage of the SUT. 

The metrics of the SUT that are collected by Prometheus are scraped from three different endpoints provided by iDRAC Exporter, cAdvisor, and Kepler. The iDRAC Exporter\footnote{\protect\url{{https://github.com/mrlhansen/idrac\_exporter}}} exposes the iDRAC metrics of the SUT. This exporter runs on the TC and interfaces with iDRAC's Redfish API to collect the required metrics. Whenever the iDRAC exporter is scraped by Prometheus, it uses the Redfish API to collect the data and converts it to the format expected by Prometheus. The power supply metrics collected from this endpoint are of particular interest, as they form the ground-truth power usage of the SUT. The node's Kubernetes metrics are made available for scraping by cAdvisor. This allows us to monitor the per-container metrics such as CPU and memory usage. Kepler additionaly exposes an endpoint for metric scraping, by default.

Finally, Kepler is deployed in the Kubernetes cluster on the SUT. Specifically, Kepler version~0.7.2 is deployed, as newer versions suffer from an issue where incorrect values are sometimes measured at random\footnote{\url{https://github.com/sustainable-computing-io/kepler/issues/1344}}. Kepler is configured to use the Redfish iDRAC integration for platform power. Since Redfish cannot provide component power, RAPL is used instead. This configuration ensures that Kepler has access to the most accurate power source, as iDRAC can measure the server power supplies directly.
A list of the metrics from each source that are used for the experiments is available in \Cref{tab:metrics}.
The Helm chart values for deploying the cluster and the observability stack are available together with the presented experimental data below in the replication package for this work\footnote{\url{https://doi.org/10.5281/zenodo.14332659}}.
%is available in the project repository\footnote{\url{https://github.com/bjornpijnacker/msc-thesis/tree/main/setup/sut/cluster/kepler}}.

% PARAGRAPH ON RAPL HAS BEEN REMOVED
% Since Kepler uses RAPL as a source of power, we also import RAPL data into Prometheus. For this purpose we write custom software\footnote{The source code is available at \url{https://github.com/bjornpijnacker/msc-thesis/tree/main/setup/sut/rapl-prometheus}} that can read the RAPL Powercap information and expose this to a Prometheus scraper. While RAPL is an Intel processor technology, the Linux power capping framework exposes the underlying RAPL information as files in the Linux filesystem \cite{noauthor_power_nodate}. For \sut, RAPL exposes---for each CPU---platform energy usage and DRAM energy usage as a counter in \unit{\micro\joule}.

\subsection{Experiment Design} \label{sec:kepler-evaluation:design}

To evaluate Kepler's container power attribution we run a simple stress workload on the test cluster, and observe Kepler's container power metrics. While running the test we investigate not only the measurements that Kepler gives for our running test container but also look at the power draw/energy usage of the other containers in the cluster. The stress workload on our Kubernetes cluster is invoked using the Linux comman \texttt{stress-ng --cpu 32 --timeout 5m}. Afterwards, we let the system sleep for 5 minutes before running the test load again. This sequence of fifteen minutes is repeated 3 times. We expect to clearly see the generated load as power usage for our test container while other containers remain stable in their power usage throughout.

Before running these tests, we also set up 16 idle containers. They will simply run the \texttt{date} command once before we start taking measurements and then exit. These containers will therefore have the `completed' status while our main workload is running. As such, they should not interfere with the dynamic power attribution by actually using power. Deploying these containers should also yield a situation closer to a real-world cluster, where a container of interest is not isolated on the cluster. However, as we do not want these containers to influence CPU utilization and power usage, they remain idle. We therefore expect these containers to not be attributed any power during the runtime of our tests.

\subsection{Results} \label{sec:kepler-evaluation:results}
\begin{figure}
    \centering
    \includegraphics[width=\linewidth]{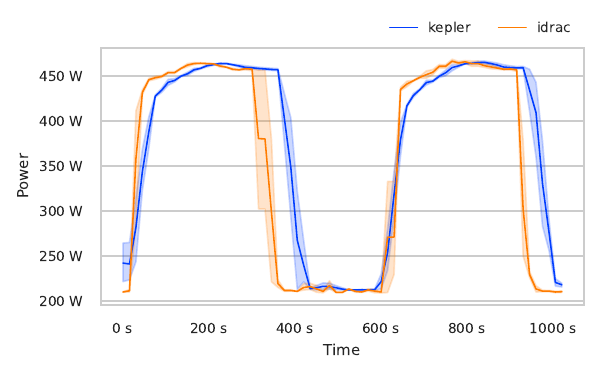}
    \caption{Total power of SUT during the stressor tests}
    \label{fig:kepler-evaluation:single-stressor--node-power}
\end{figure}

\begin{figure}
    \centering
    \includegraphics[width=\linewidth]{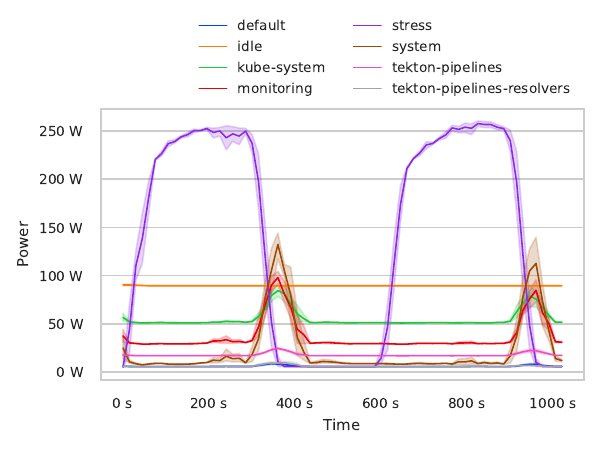}
    \caption{Per-namespace container power draw as reported by Kepler during the stressor tests}
    \label{fig:kepler-evaluation:single-stressor--namespace-power}
\end{figure}

\Cref{fig:kepler-evaluation:single-stressor--node-power,fig:kepler-evaluation:single-stressor--namespace-power} visualize the test results for our Kepler stressor test. In \cref{fig:kepler-evaluation:single-stressor--node-power}, the total power for SUT as measured by both iDRAC and Kepler is indicated. Visually, the measurements align quite closely. Between iDRAC and Kepler there is a root mean square error (RMSE) of \qty{66.4}{\watt}. This rather large error is mostly caused by differences in reporting latency between iDRAC and Kepler, as also evident in the figure. Considering the total energy usage during the test instead of the wattage over time, we see an error of less than \qty{1}{\percent}, which is more than acceptable.

\begin{figure}
    \centering
    \includegraphics[width=\linewidth]{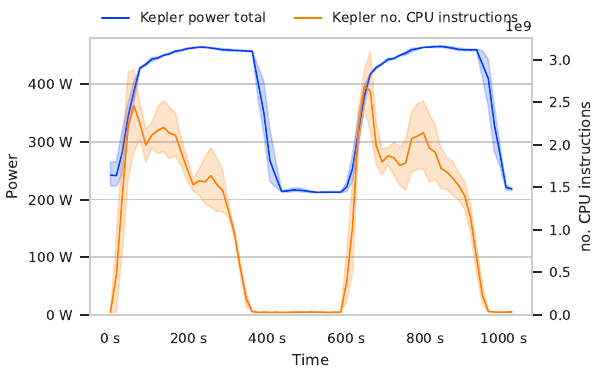}
    \caption{Power usage and number of CPU instructions as reported by Kepler during the stressor tests}
    \label{fig:kepler-evaluation:single-stressor--cpu-timing}
\end{figure}

\Cref{fig:kepler-evaluation:single-stressor--namespace-power} shows the attributed container power of Kepler summed by Kubernetes namespace. The results are partially as expected: the power attributed to the `stress' namespace, which houses exclusively the stress-test container running our workload, corresponds to the generated load which was expected. Two things in this figure are not as expected however:
\begin{enumerate}
    \item there is a peak of power usage in other namespaces after the stress-test stops, and
    \item the `idle' namespace consistently uses approximately \qty{100}{\watt}, even though it should not be using any energy.
\end{enumerate}
The peak can be explained by considering that iDRAC may measure power usage slower than Kepler measures CPU usage. We verify this in \cref{fig:kepler-evaluation:single-stressor--cpu-timing}, where we indeed see that power consumption metrics lag behind CPU load metrics by up to \qty{1}{\minute} when CPU load quickly decreases. This finding is in-line with the specification provided by iDRAC and Redfish, as this specifies power metrics\footnote{Available through the \url{redfish/v1/Chassis/System.Embedded.1/Power/PowerControl} API endpoint.} are updated on a one-minute interval. In this situation, where not all metrics are updated at the same interval, Kepler needs to attribute more power usage among containers than is actually occurring at that given moment, therefore artificially spiking the power usage of all workloads as CPU usage suddenly decreases.

We additionally see unexpected power usage of the `idle' namespace, with all containers in the `idle' namespace as reported to be using around \qty{6.25}{\watt} each. Recall that we created these containers as part of the test, and that the containers are not running. All power attribution to these containers is of the idle power mode, where the power attributed to our stress container was dynamic. This means that Kepler is effectively reporting power usage for containers that are not actually running. While there is indeed overhead to managing non-running containers in Kubernetes, idle-mode power should not be equally divided among them, as this may lead to unfair attribution when the number of containers on the cluster changes. Since the total power usage of all containers was correct with respect to the iDRAC measurement, this also indicates that Kepler is at the same time likely under-reporting the idle power used by the other containers to begin with.

To see whether Kepler can correctly attribute power once the inactive containers have been deleted, we perform the following test. First, we start 64 idle containers that run \texttt{date} and exit as before. We choose a large number of idle containers so that their presence and absence has a large and thus easy to observe effect on measurements. After these pods are created and complete we give the system one minute to stabilize. We then run a small (8 CPU) stressor and delete all idle containers after two minutes, then observe how the container attribution of Kepler changes. We expect that Kepler reallocates the idle power usage to all other containers, and that the dynamic power attribution does not change.

\begin{figure*}
    \centering
    \subcaptionbox%
        {Per-namespace container power\label{fig:kepler-evaluation:inactive-pods--namespace-total}}%
        {\includegraphics[width=0.49\linewidth]{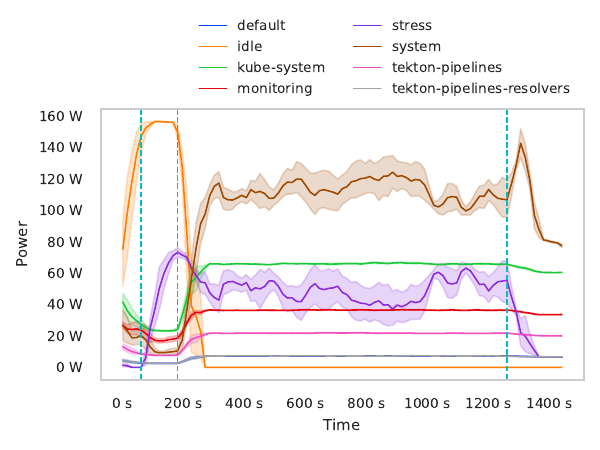}}
    \subcaptionbox%
        {Per-namespace idle-mode container power\label{fig:kepler-evaluation:inactive-pods--namespace-idle}}%
        {\includegraphics[width=0.49\linewidth]{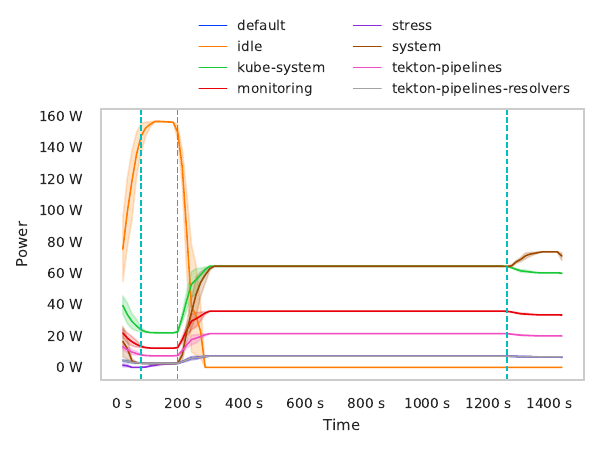}}
    \subcaptionbox%
        {Per-namespace dynamic-mode container power\label{fig:kepler-evaluation:inactive-pods--namespace-dynamic}}%
        {\includegraphics[width=0.49\linewidth]{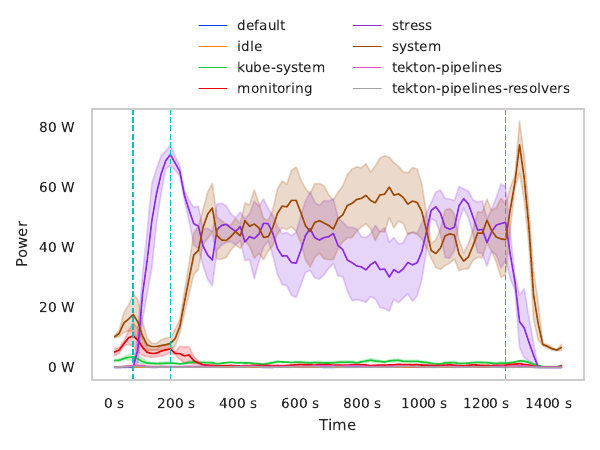}}
    \subcaptionbox%
        {Per-namespace CPU utilization as reported by cAdvisor\label{fig:kepler-evaluation:inactive-pods--cpu-utilization}}%
        {\includegraphics[width=0.49\linewidth]{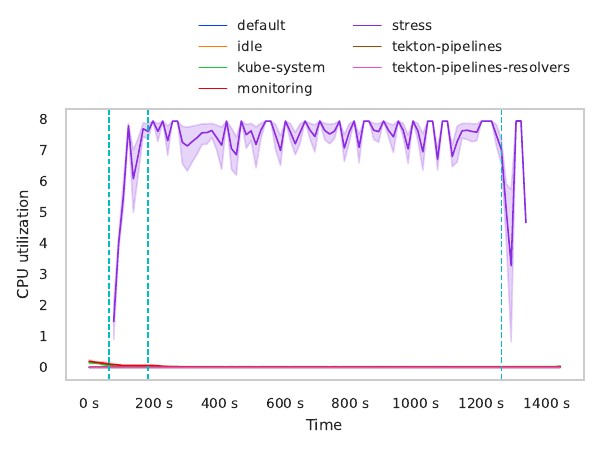}}
    \caption[foo]{Power attribution and CPU utilization before and after deleting inactive pods. The markers indicate the times at which \begin{enumerate*}[label=(\arabic*)]
            \item the stressing load was started;
            \item the inactive pods were deleted;
            \item the stressing load was stopped,
        \end{enumerate*} respectively.}
    \label{fig:kepler-evaluation:inactive-pods}
\end{figure*}

The results of this test are presented in \cref{fig:kepler-evaluation:inactive-pods}. The test was repeated four times. \Cref{fig:kepler-evaluation:inactive-pods--namespace-total} shows the power attributed by Kepler to each namespace. \Cref{fig:kepler-evaluation:inactive-pods--namespace-idle,fig:kepler-evaluation:inactive-pods--namespace-dynamic} show this for idle- and dynamic-mode power respectively. Recall that the `stress' namespace contains solely our stressor pod and that the `idle' namespace only has the 64 inactive pods. As the test starts, the total power goes up for the `stress' namespace as expected. As the idle pods are deleted, the idle-mode power for the `idle' namespace quickly goes to zero as expected. Power is re-attributed throughout the other namespaces over all remaining containers.

After deleting the idle pods we also see the dynamic power usage for the `stress' namespace going down and the dynamic power usage for the `system' one go up. Note that the CPU usage of the workload remained at 100\%, and consistent throughput was indicated throughout the experiment as per the stressor logging, as also indicated in \cref{fig:kepler-evaluation:inactive-pods--cpu-utilization}. The change in power attribution here is unexpected, since there is no `system' namespace with running workloads. The upward trend of this namespace's power attribution matches a downward trend in reported dynamic power usage for the `stress' namespace. According to Kepler, this power is being attributed to pods named `system\_processes', which is a reserved name in Kepler for processes that cannot be attributed to a pod. After stopping the testing workload running in the `stress' namespace, we see that the `system' namespace also reports using less power. %This indicates very clearly that Kepler is not able to properly attribute the power used by the stressing container to that container.

\subsection{Interpretation} \label{sec:kepler-evaluation:interpretation}
The container power attribution in Kepler leaves much to be desired. In our experiments, we have seen multiple examples of container power attribution that were clearly inaccurate and sometimes inexplicable, both in idle- and dynamic-power mode. As we have seen in the inactive-container deletion test, Kepler misattributes some power to system processes, where it should not be doing so. Additionally, as we have seen in the single-stressor test, Kepler attributes power to non-running containers, which is also not correct. Consequently, and as a response to RQ1, \emph{Kepler does not seem to produce a trustworty measure of container energy usage in Kubernetes clusters}.
The scripts to run and the resulting datasets from these experiments are available in the replication package for this work.
%\footnote{Double blind review-friendly \textbf{\href{https://zenodo.org/records/14332660?token=eyJhbGciOiJIUzUxMiJ9.eyJpZCI6IjRkOWEyMzc5LWQ0ZjItNGU4MS05OTY1LTc0MjE0Njg2NjRkMiIsImRhdGEiOnt9LCJyYW5kb20iOiIxNTM4ZDU1YWMxMzAyODc3Mjc4YjMyMjcyNmUxNmVhZiJ9.9J-7B7UL-2QYLsSGZJl-pPLhAneeoAbu7c5q8OkEjPDr3dpBbve1n3O4EyGbGA7mhmvXkEg21KwXhd48jLrq1Q}{link to zenodo}}.}.
%\footnote{\url{https://doi.org/10.5281/zenodo.14332660}}.

% \todo[inline]{@Bjorn: I was going to write on Kepler's custom model here except we are not evaluating node measurements here. It doesn't make sense to me to place it here. Note to self: talk about why the power source does or does not matter (relatively) when considering only container attribution.}

\section{KubeWatt} \label{sec:kubewatt}
In the previous section, we have shown that Kepler is not a suitable tool for producing container-level power metrics in Kubernetes with sufficient accuracy. As an alternative, we create our own tool: KubeWatt. It will be based partially on the power attribution model which is proposed in \cite{andringa_estimating_2024}, since the latter yielded promising results in container/pod power attribution.

Functionally, KubeWatt can read CPU utilization metrics from Kubernetes, obtain node power usage using a Redfish API integration, can split node power usage into static and dynamic parts and can produce Prometheus-style metrics for container power usage. The first two are obvious: in order to calculate power usage of a container, KubeWatt must at least know the total power usage of the node that container is running on, and it must know the amount of resources that container uses on the node. Dividing the power usage into `static' (or `idle' in Kepler) and `dynamic' parts aims to account for the power usage of a Kubernetes cluster and server when no workloads are running. The simple act of turning on a server and running a Kubernetes cluster uses some amount of power which cannot be attributed to any specific container. KubeWatt therefore splits the total power into two components, such that the static power can be indicated in total, and the dynamic power, which is the difference between static and total power, can be attributed amongst Kubernetes containers. KubeWatt will indicate static power as a single number in its output, since it indicates overhead that is not easily attributed to any specific container. Note that this definition of overhead includes Kubernetes control plane containers. These are therefore excluded from the dynamic power attribution. Any additional power usage incurred by the control plane will instead be attributed to other running containers causing the control plane to use power.

\subsection{Allocation Model} \label{sec:kubewatt:allocation-model}
KubeWatt builds on the allocation model proposed in \cite{andringa_estimating_2024} to attribute total power among containers in Kubernetes, named `pod mapping'. We do, however, make a few changes to fit the model to our purpose.
More specifically, we define \( \textrm{power}(\cdot) \) as the power usage of some component and \( \textrm{cpu}(\cdot) \) as the CPU utilization of some component. We define \( \textrm{power}_d(\cdot) \) and \( \textrm{power}_s(\cdot) \) as the dynamic and static fractions of power as described above, respectively, with \[ \textrm{power}(\cdot) = \textrm{power}_d(\cdot) + \textrm{power}_s(\cdot). \] Then let \( n_i \) be some Kubernetes node whose power \( \textrm{power}_d(n_i) \) and \( \textrm{power}_s(n_i) \) are known. Let \( c_{m,i} \) be a Kubernetes container running on node \( n_i \), and identified uniquely by \( m \). Given the CPU utilization of \( c_{m,i} \), we have
\begin{equation} \label{eq:kubewatt:allocation-model:container-power}
    \textrm{power}(c_{m,i}) = \textrm{power}_d(n_i) \cdot \frac{\textrm{cpu}(c_{m,i})}{\sum_i \textrm{cpu}(c_{m,i})}.
\end{equation} Importantly, we make a distinction between \( \sum_i \textrm{cpu}(c_{m,i}) \), the CPU usage of all containers on a node combined, and \( \textrm{cpu}(n_i) \) the CPU utilization of the node, as the metrics API of Kubernetes includes overhead CPU usage such as system processes in the latter metric, which is already included in static power and should not be attributed to Kubernetes containers\footnote{https://kubernetes.io/docs/reference/external-api/metrics.v1beta1/}. \Cref{eq:kubewatt:allocation-model:container-power} gives us a metric of power usage for each container in a Kubernetes node as derived from the node power usage and container CPU utilization.

\subsection{Architecture} \label{sec:kubewatt:architecture}
\begin{figure}[t]
    \centering
    \includegraphics[width=0.7\linewidth]{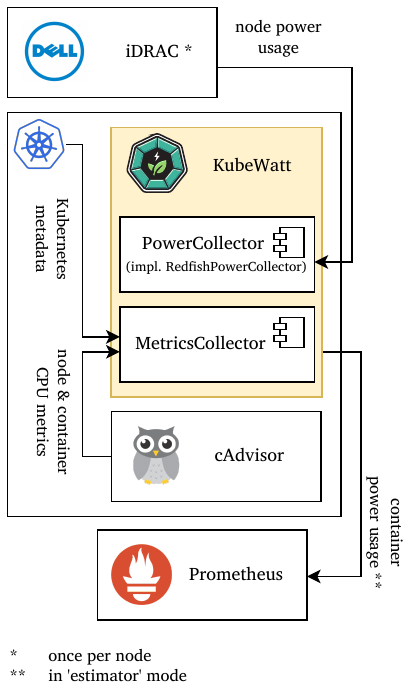}
    \caption{High-level diagram of KubeWatt architecture}
    \label{fig:kubewatt:architecture}
\end{figure}

KubeWatt consists of several components which interface with other applications. In this section, we discuss the \textit{power collector} and \textit{Kubernetes metrics collector} components. A high-level diagram of KubeWatt in context is shown in \cref{fig:kubewatt:architecture}.
More specifically:

\subsubsection*{Power collector}
The power collector component of KubeWatt is responsible for providing a measure of power usage in \si{\watt} per node in the Kubernetes cluster. KubeWatt does not care about the source of power and the implementation is abstracted behind the \texttt{PowerCollector} interface, meaning that it is easily extended to use other power sources. For the purposes of this work, the only implemented version of this interface is the \texttt{RedfishPowerCollector} class, which uses the Redfish API in SUT to obtain power usage from the power supply. 

% The RedfishPowerCollector collects power information from server management interfaces which implement the Redfish API. For each Kubernetes node named in config field \texttt{collector.node-names}, KubeWatt expects a corresponding Redfish entry in the \texttt{collector.power.redfish} map. Each of those must contain a host, username, password and list of Redfish ComputerSystem names. Since a single Redfish API can return multiple systems \autocite{RedfishDataModel}, we require that the user of KubeWatt specify which system(s) correspond to which Kubernetes node. The power readings for each system are summed per Kubernetes node to provide the final power values. 

\subsubsection*{Kubernetes metrics collector}
The Kubernetes metrics collection component is responsible for obtaining CPU usage metrics of both nodes and pods in the Kubernetes cluster. It uses the \texttt{metrics.k8s.io/v1beta1/pods} and \texttt{metrics.k8s.io/v1beta1/nodes} Kubernetes API endpoints for pods and nodes metrics respectively. The nodes endpoint returns CPU utilization in \unit{\nano\second} for each node. The pods endpoint returns CPU utilization in \unit{\nano\second} for each container in each pod\footnote{https://kubernetes.io/docs/reference/external-api/metrics.v1beta1/}. In the Kubernetes Java API implementation, which KubeWatt uses, the pods metrics are available per namespace\footnote{https://github.com/kubernetes-client/java}.

%\subsubsection{Initialization process}
\subsection{Operational modes}
KubeWatt runs in one of three modes. The first two modes, `base initialization' and `bootstrap initialization' are initialization modes. They run as a one-off job to initialize KubeWatt parameters. These modes analyze the cluster which KubeWatt will run on and find the static power value (\( \textrm{power}_s \)) for each node in the cluster. This value is calculated once and not updated without manual intervention; therefore, we assume that this does not change over time.

% In my thesis I show that this is a relatively safe assumption to make but I skip that tangent here.
\subsubsection{Initialization modes}
The \textit{base initialization mode} is the simplest mode that KubeWatt can run in. It expects an empty Kubernetes cluster running no more than the Kubernetes control plane. KubeWatt expects its user to specify which pod names are part of the control plane as a set of regular expressions, such that it can validate the cluster is indeed empty before starting. Over a period of 5 minutes, KubeWatt measures the power usage per node every fifteen seconds, which is averaged to produce the static power value per node. We expect that this mode will produce the most accurate results for the static power value, as it directly measures the idle cluster.

The \textit{bootstrap initialization mode} is an alternative to the base initialization mode. It attempts to find the static power value from a cluster which is already running workloads that cannot be turned off for testing. Since this mode makes an estimation of the static power value based on measurements, the base initialization mode should be preferred if possible. This mode gathers both CPU-usage data for each node and power usage data for each cluster node. Data are gathered every fifteen seconds for half an hour. Afterwards, KubeWatt checks whether the data have enough variability and a sufficient distribution to draw conclusions from. If not, data collection is repeated, otherwise KubeWatt proceeds with data analysis.
To gauge whether data are sufficient to perform analysis with, KubeWatt checks the rough distribution of the data. The collected CPU-usage values are placed in buckets. KubeWatt then checks the amount of measurements in the largest bucket, and validates that no bucket has fewer values than some predefined factor of it. By default, each bucket should have at least half as many values as the largest, to ensure a relatively uniform distribution. The buckets are \qty{10}{\percent} in size between \qty{20}{\percent} and \qty{80}{\percent} CPU utilization by default. All of these values are configurable. 
% Changing the bucket parameters may be useful in a cluster which never reaches as low as \qty{20}{\percent} CPU. This does not hold for the upper bound, since the user can run an artificial stressor to generate higher CPU-load.

As previously discussed, the static power should include the Kubernetes control plane utilization at idle. To achieve this, the CPU-usage of the control plane containers is gathered at the same time as the node CPU-usage and power usage. The set of control plane containers is known as the user is required to specify these when configuring KubeWatt. The control plane CPU utilization is averaged to give an indication of stable control plane CPU-usage at idle. Note that we cannot expect the control plane CPU utilization to remain stable when the cluster has load. As the idle control plane utilization is encapsulated in the static power value, any power usage caused by higher utilization in result of cluster load will be attributed among the containers causing this load.
To finally derive the static power usage, a linear regression is performed on the collected data below \qty{50}{\percent} CPU utilization, as we know from \cite{kistowski_analysis_2015} that CPU utilization against power usage grows linearly for those values. The regression coefficients are subsequently used to find the estimated power usage at the average CPU-usage of the control plane. This then gives us the static power for each node in the Kubernetes cluster.

\subsubsection{Estimation mode}
The estimation mode is what we consider the `main' mode of KubeWatt. This mode takes the output from either initialization mode as input and actually estimates the amount of power that each container in the Kubernetes cluster uses. When running in this mode, KubeWatt exports Prometheus-style metrics. The estimator mode uses the allocation model as described in \Cref{sec:kubewatt:allocation-model}.

\subsection{Deployment} \label{sec:kubewatt:deployment}
KubeWatt can be deployed in a Kubernetes cluster using the provided Helm chart\footnote{\url{https://github.com/bjornpijnacker/kubewatt}}. There are two important configuration options that need to be set, the Redfish configuration mapping and the operation mode. It is important that the Redfish REST interface is accessible from the node on which KubeWatt is deployed. The exact configuration required to setup Redfish is dependent on the server manufacturers out-of-band management software. 
It is important to ensure that KubeWatt is configured to first run in one of the two initialization modes \texttt{INIT\_BASE} or \texttt{INIT\_BOOTSTRAP}. The base initialization mode should only be used on an empty cluster, whereas the bootstrap initialization mode should be used on an already provisioned cluster. The initialization modes will output the static power values per node. This static power value should then be configured in the value file, after which KubeWatt can be run in the \texttt{ESTIMATOR} mode to begin estimating the power usage per Kubernetes container. 

\section{KubeWatt Evaluation} \label{sec:kubewatt-evaluation}
In the previous section, we have outlined how KubeWatt works. In this section, we perform an evalution on KubeWatt in order to validate its workings and to showcase whether or not we have improved on Kepler's limitations as discussed in \Cref{sec:kepler-evaluation}. For this purpose we pose three research questions:
% \todo[inline]{Define RQ(s); kwRQ3, kwRQ4; kwRQ2 -> RQ2 RQ3 RQ4 (in that order!)}
\begin{tcolorbox}[left=3pt, right=3pt, top=3pt, bottom=3pt]
\begin{enumerate}[label=\textbf{RQ\arabic*},left=0pt]
    \setcounter{enumi}{1}
    \item How accurately can KubeWatt's base initialization mode report the static power value?
    \item How accurately can KubeWatt's bootstrap initialization mode estimate the static power value?
    \item How (well) does KubeWatt attribute power usage to containers on a Kubernetes node?
\end{enumerate}
\end{tcolorbox}

The testing setup as discussed in \Cref{fig:kepler-evaluation:system-under-test:setup} is reused as-is; however, the Kepler deployment has been removed as this is no longer necessary. KubeWatt is now deployed in its place.

For the initialization modes, we want to verify that the output that KubeWatt gives is equal to the static power of the system. The static power value has been determined from iDRAC data at \qty{199.1}{\watt}. To evaluate the base initialization mode (RQ2), we run the KubeWatt job on SUT with nothing else running on the cluster. The following list of pod names is provided to KubeWatt as control plane: \texttt{nfs-.*}, \texttt{calico-.*}, \texttt{canal-.*}, \texttt{coredns-.*}, \texttt{metrics-.*}, \texttt{tekton-.*}, \texttt{kubewatt-.*}. Note that \texttt{tekton-.*} is not technically part of the control plane; however, it could not be easily removed and as it is idle, it should not have a significant impact on findings. We repeat the job six times in sequence, then verify that the output value closely corresponds to the expected power value. To obtain the expected power value, the power use reported by iDRAC is tracked during the runtime of each of the initialization job runs. For the base initialization mode we evaluate whether the resulting values are consistent over multiple runs and whether the resulting values are accurate to the expected value taken from iDRAC.

To evaluate the bootstrap evaluation (RQ3), we run a best-case test, where the cluster is stressed with a random stressor. Stress-ng is used to create stressors at a random CPU level between 1--64 that last three minutes each. This creates a CPU load that should be uniformly distributed across the entire CPU range of SUT. This test is repeated three times, again ensuring consistency and accuracy.
% Since it is not possible to use a synthetic load to \emph{decrease} the CPU utilization below what a running system normally has, KubeWatt allows modification of the minimum CPU value it checks for when validating the data. Tuning this value avoids a situation where KubeWatt's bootstrap initialization never finishes due to insufficient data in a bucket which will never get any data. Note that KubeWatt does \emph{use} the data outside of buckets, it only does not validate that it exists prior to continuing with the analysis. To show the effect a limited range of data has on the output of KubeWatt we make cuts of the dataset obtained by the earlier tests, then run the regression as KubeWatt would to observe how the output would change.
% THIS PART IS SKIPPED
%
To evaluate the estimation mode we repeat the test we ran for Kepler in which we deleted inactive pods while running a stressor. This test is meant to validate both that KubeWatt can accurately gauge container power usage as well as showcase that it does not suffer from the same limitations as Kepler.
The replication package for this work contains all the necessary scripts and datasets, as discussed in Section~\ref{sec:kepler-evaluation}.

\subsection{Results}  \label{sec:kubewatt-evaluation:results}
\subsubsection{Base initialization mode}
Running the base initialization on our empty cluster, KubeWatt reports static power values of \qty{198.9}{\watt}, \qty{199.15}{\watt}, \qty{199.1}{\watt}, \qty{199.1}{\watt}, \qty{198.75}{\watt}, and \qty{199.15}{\watt} across the six tests. These values are within \qty{0.35}{\watt} of each other and within \qty{0.3}{\watt} of the expected value, or in other words within an error of less than \qty{0.2}{\percent}, indicating a consistent and accurate output.

\subsubsection{Bootstrap initialization mode}
\begin{figure*}
    \centering
    \includegraphics[width=\linewidth]{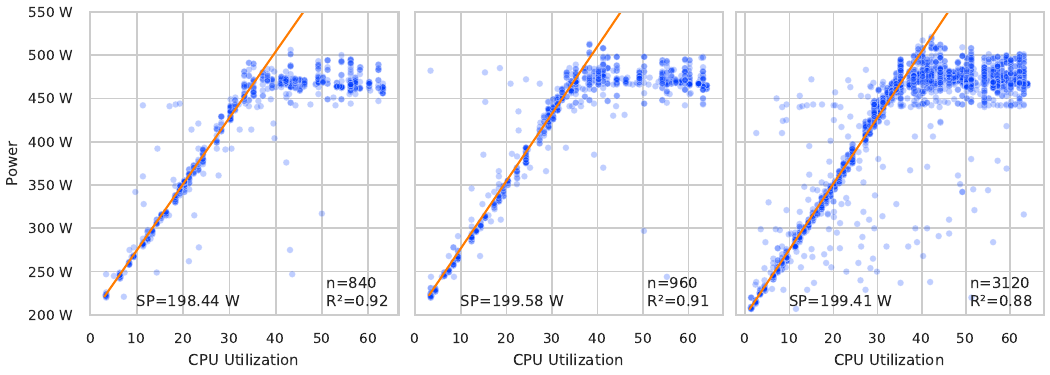}
    \caption{CPU utilization and power usage (blue) as reported by KubeWatt for the bootstrap initialization mode. Result of linear regression on bottom half of data (orange). Repeated three times.}
    \label{fig:kubewatt-evaluation:results:bootstrap-initialization}
\end{figure*}

The measurements that KubeWatt has taken for the bootstrap initialization mode are shown in \cref{fig:kubewatt-evaluation:results:bootstrap-initialization} for each of the three repeated tests. The raw values are shown in blue, while the resulting regression that KubeWatt performs is shown in orange. Recall that KubeWatt only performs the regression on the lower \qty{50}{\percent} of the data. %to account for the knee that is present in the measurements.
KubeWatt reports static power values of \qty{198.44}{\watt}, \qty{199.58}{\watt}, and \qty{199.41}{\watt}, again indicating consistent and accurate output. For each of the tests, the control plane utilization contributed approximately \qty{0.30}{\watt}. It is noteworthy that these tests, and the bootstrap initialization mode in general, take a lot longer to run than the base initialization mode. The three tests as shown in \cref{fig:kubewatt-evaluation:results:bootstrap-initialization} took, respectively, \qty{3.5}{\hour}, \qty{4}{\hour} and \qty{13}{\hour} to finish.

\subsubsection{Estimation mode}
\begin{figure}[t]
    \centering
    \subcaptionbox%
        {Per-namespace container power\label{fig:kubewatt-evaluation:results:estimation--namespace}}%
        {\includegraphics[width=\linewidth]{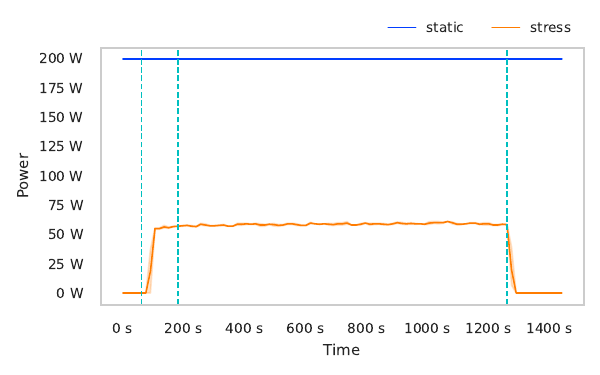}}
        
    \subcaptionbox%
        {Per-namespace CPU utilization as reported by cAdvisor\label{fig:kubewatt-evaluation:results:estimation--cpu-utilization}}%
        {\includegraphics[width=\linewidth]{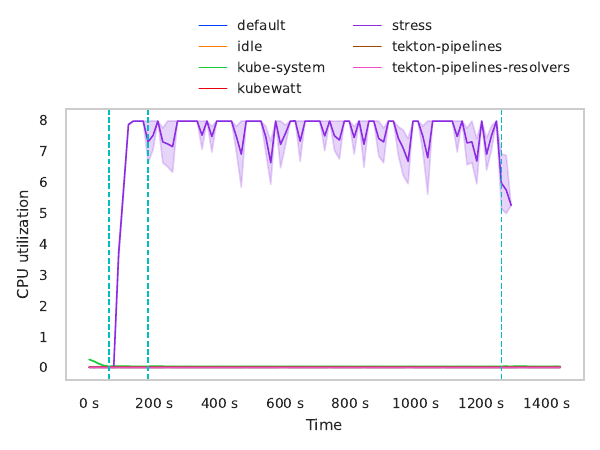}}
    \caption[foo]{Power attribution reported by KubeWatt and CPU utilization when running `deleting inactive pods' test. The markers indicate the times at which \begin{enumerate*}[label=(\arabic*)]
            \item the stressing load was started;
            \item the inactive pods were deleted;
            \item the stressing load was stopped,
        \end{enumerate*} respectively.}
    \label{fig:kubewatt-evaluation:results:estimation}
\end{figure}

The results of the repeated `deleting inactive pods' test is shown in \cref{fig:kubewatt-evaluation:results:estimation}. \Cref{fig:kubewatt-evaluation:results:estimation--namespace} shows the per-namespace container power. Note that `static' is not a namespace, but is the namespace-less static power. Some namespaces here are missing compared to \cref{fig:kepler-evaluation:inactive-pods}; this is because all containers in those namespaces are part of the control plane and are not reported separately by KubeWatt. Note also that the idle containers are not reported. This is the case because they are not using power: they are not running. \Cref{fig:kubewatt-evaluation:results:estimation--cpu-utilization} shows the per-namespace CPU utilization. We see the same measurements as in \cref{fig:kepler-evaluation:inactive-pods--cpu-utilization}, indicating the tests were performed equally.

The main, and expected result is depicted in \cref{fig:kubewatt-evaluation:results:estimation--namespace}. We see that the power usage of the stressor pod goes up with CPU utilization, stays consistent throughout the test and goes down when the stressor is stopped. Conclusively, we see from these figures that KubeWatt can accurately report the power usage of the stress pod, even when a large number of inactive pods clutter the cluster.

\subsection{Discussion}  \label{sec:kubewatt-evaluation:discussion}
With the results of KubeWatt experiments that have been outlined, we can now turn again to the questions posed at the beginning of this section.
The base initialization mode is able to very accurately read the static power value (RQ2). In our test we have seen the base initialization mode get very consistent measurements over time, with the largest internal difference being \qty{0.3}{\watt} out of a \qty{199.1}{\watt} mean over six KubeWatt runs. Additionally, we have seen that the bootstrap initialization mode can also accurately estimate the static power value (RQ3). In this case, KubeWatt can estimate the static power value within \qty{0.7}{\watt} of the expected value in a best-case scenario.
% , and can estimate the static power value within \qty{5}{\percent} of the expected value with data only above \qty{40}{\percent} CPU utilization.
In terms of container power attribution (RQ4), we see that KubeWatt is able to accurately portray the power used by a container based on CPU load, in line with what is expected based on CPU utilization, while being consistent with the total node power draw ground-truth measurements.

% "KubeWatt does incorrectly attribute power when containers are newly created and immediately generating load when other containers already exist." is in my thesis but that test is skipped here. This is the single-stressor test which we didn't repeat for KubeWatt in this paper.

\subsection{Limitations}  \label{sec:limitations}
% In this section, we consider threats to validity of this research project.
There are some obvious limitations to the evaluation of KubeWatt, effectively constituting threats to the validity of this procedure, and we discuss them in the following.

% RAM errors
%The main server, SUT, that was used for running tests was not fully functional. One of the twelve memory sticks raised bit errors occasionally: ``the system memory has faced an uncorrectable multi-bit memory errors in the non-execution path of a memory device''. Test results were discarded when an error occurred during the test; however, the overhead of performing error correction may cause more noisy CPU load and thus influence results and test repeatability.

% only a single server and non-ideal deployment
First, the SUT comprised a single server only. Evaluation results would have been more robust by combining a number of servers and distributing multiple workloads among them to more closely resemble a real-world cloud environment.
KubeWatt has been designed and implemented with the general case in mind, but lack of access to multiple server devices and of time to conduct a full blown experiment prohibited us from demonstrating this; it is left as future work.
%
% The server in question was not deployed in an ideal scenario. Our server was not deployed in a real datacenter due to the need for management port (iDRAC) access which the University of Groningen would not allow in their datacenters. The room where the server was installed had a temperature which fluctuated between \qty{25}{\degreeCelsius} and \qty{28}{\degreeCelsius} during testing which may cause fluctuation in total power-usage and influence repeatability of tests.
%
% only fake workloads
Furthermore, during testing, we only ran artificial workloads using \texttt{stress-ng}. While these stress the CPU, they do so in a very specific way which may not always be equal to a real-world workload. This means that the shown results may not be completely indicative of real-world behavior.
However, we have no reason to believe that the accuracy of our results would not replicate also in other applications since we make no assumption concerning the workload itself in KubeWatt.

Last but not least, our KubeWatt implementation has only been evaluated in what is for all practical purposes a private cloud deployment.
Moving, for example, to a public cloud deployment would mean that we do not have access anymore to iDRAC (or equivalent) data through the Redfish API and we would also have to rely as Kepler does in this case on either RAPL data where available, or on an estimator model.
Demonstrating that KubeWatt can produce consistent, if not accurate results in this scenario is left as future work.
For the purposes of this experimental procedure, however, both Kepler and KubeWatt use iDRAC, which allows them to be compared reliably.
An important point concerning also the evaluation of Kepler as discussed in \Cref{sec:kepler-evaluation} was not following the option of developing a custom power model for the latter.
Our experience in attempting to do so was frankly frustrating to disappointing, with large gaps left in the documentation of the task that we were not able to cover on our own. 
Being able to build a custom power model could have perhaps resulted into more favorable for Kepler results.

\section{Related Work} \label{sec:related-work}
Existing approaches in the literature have looked into providing observability for cloud-native applications but outside of Kubernetes and Kepler. Fieni et al. \cite{fieni_smartwatts_2020}, for example, introduce \textsc{SmartWatts}, a tool for estimating container energy consumption. It works by considering component energy and hardware performance counter events to gauge resource utilization. Their model then maps power usage into static and non-static power, and divides this over the resource utilization. \textsc{SmartWatts} uses Linux Cgroups to allow a wide range of monitoring granularities, such as Docker containers, Virtual Machines or processes. % Intel RAPL is used through MSR (Model Specific Registers) to obtain power readings. The authors name that RAPL is not perfect for the application, nor are other power meters such as IMPI. For example, RAPL is only available on bare-metal environments while IMPI is not sufficiently granular. To this end, 
In a follow up work, the authors propose \textsc{SelfWatts} \cite{fieniSelfWattsOntheflySelection2021}, as a novel method to estimate component power utilization. %\textsc{SmartWatts} is shown to have a \qty{2}{\watt}--\qty{3}{\watt} error for container power estimation in performed tests, and \textsc{SelfWatts} is shown to have an at-most \qty{2}{\watt} error in VM power estimation.

Dinga et al. \cite{dingaEmpiricalEvaluationEnergy2023} examine the performance overhead of monitoring solutions on Docker-based systems. The monitoring applications under test were ELK Stack, Netdata, Prometheus and Zipkin. They showcase a significant effect on power consumption for several of the tools and scenarios, with a \qty{1.47}{\percent} to \qty{12.86}{\percent} increase in energy consumption depending on the tool. They furthermore show a significant impact on CPU usage for the ELK Stack and Zipkin, and in RAM usage for all tools except Netdata.
Santos et al. \cite{santosHowDoesDocker2018} investigate the impact of running an application in Docker versus running it on bare-metal Linux. A power meter is used to track the total system power usage when running the tests. The authors find that having the Docker service \texttt{dockerd} running the background uses a significant amount of power above bare-metal idle. %For the Redis and WordPress benchmarks, the authors find that the runtime of the benchmark in increases, and the energy consumption increases correspondingly, when running in Docker.
In contrast to those works, our approach focuses on workloads specifically deployed in Kubernetes clusters.
On the other hand, works that do focus on Kubernetes, such as those discussed in Section~\ref{sec:kepler}, lack the necessary granularity and depth in their assessment --- a gap that this work bridges as discussed in the previous sections.

%Vitali et al. \cite{vitaliEnrichingCloudnativeApplications2023} also consider the application-perspective of cloud-native application sustainability. They introduce a design architecture that enriches cloud-native applications with sustainability features. The architecture is characterized by three subsystems; one for the application designer, one for the application developer and one for the infrastructure provider.
% Removed a snippet to shorten related work. For reference:
% In the proposed process, an application microservice architecture is represented through Business Process Model and Notation, which is extended to include cloud native sustainability task types. This allows specifying specific tasks as optional or mandatory or to define multiple variants for each task. This can aid in producing a sustainable application architecture or evaluating the sustainability of an existing architecture.
%While the authors name that sustainability benefits of their architectural components cannot be quantified at the current stage, they do show a \qty{27}{\percent} decrease in computational resource demand for one of their demonstrative applications.

\section{Conclusions} \label{sec:conclusions}

Through a set of controlled experiments, in the previous sections we have shown that the state-of-the-art energy measuring tool for Kubernetes clusters, Kepler, is able to produce accurate results at cluster- but not individual-container levels.
These inaccuracies appear to be fundamental to the power allocation model used by Kepler, and can result into serious misalignment if used e.g. to attribute carbon footprint among different workloads (applications) running on a cluster.
As a response to this issue, we developed KubeWatt and demonstrated that for the same scenarios it can measure power draw at both node and container levels within statistical error margins.
More importantly, the power allocation model built in KubeWatt appears to produce more internally consistent results than the one used by Kepler.

Future work aims to address the limitations identified in Section~\ref{sec:kubewatt-evaluation}, namely, expanding the experimental evaluation in the generic case of a heterogeneous Kubernetes cluster using a set of real-world application workloads. 
Furthermore, by adding carbon intensity stream data to KubeWatt, we can also offer carbon footprint measurements at different granularity levels to complement the energy measurement ones.

%\section*{Acknowledgements}
%\todo[inline]{add project acks}

\bibliographystyle{IEEEtran}
\bibliography{IEEEabrv,main}

\end{document}